\documentclass[12pt]{iopart}

\usepackage{graphicx}
\usepackage{dcolumn}
\usepackage{bm}

\usepackage{verbatim,vmargin,setstack,calc, mathrsfs}
\usepackage{amsthm, amssymb, dsfont}
\usepackage{epsfig, pstricks, pst-plot,pst-node}
\usepackage{graphicx}
\usepackage{psfrag, wasysym}

\newpsobject{showgrid}{psgrid}{subgriddiv=10}

\newcommand{\op}[1]{\mathsf{#1}}
\newcommand{\bra}[1]{\langle #1 \vert}
\newcommand{\ket}[1]{\vert #1 \rangle}
\newcommand{\scal}[2]{\langle #1 \vert #2 \rangle}
\newcommand{\mom}[2]{\langle #1^{#2} \rangle}
\newcommand{\av}[1]{\langle #1 \rangle}
\newcommand{\Id}{\mathds{1}}

\newcommand{\fracpart}[1]{\frac{\partial}{\partial #1}}
\newcommand{\fracpartat}[3]{\left. \frac{\partial}{\partial #1}{#3} \right|_{{#1}={#2}}}
\newcommand{\fracpartn}[2]{\frac{\partial^{#2}}{\partial {#1}^{#2}}}

\newcommand{\Ai}{\text{Ai}}

\newcommand{\unit}[1]{\mathsf{#1}}
\begin{document}

\title[1d Directed Sandpile Models and the Area under a Brownian Curve]
{One-Dimensional Directed Sandpile Models and the Area under a Brownian Curve}

\author{M A Stapleton}
\address{Blackett Laboratory, Imperial College London,
Prince Consort Road, London SW7 2BW, United Kingdom}
\ead{ms599@imperial.ac.uk}
\author{K Christensen}
\address{Blackett Laboratory, Imperial College London,
Prince Consort Road, London SW7 2BW, United Kingdom}
\ead{k.christensen@imperial.ac.uk}

\date{\today}

\begin{abstract}
We derive the steady state properties of a general directed ``sandpile''  model in one dimension.
Using a central limit theorem for dependent random variables we find the precise
conditions for the model to belong to the universality class of the Totally Asymmetric
Oslo model, thereby identifying a large universality class of
directed sandpiles.
We map the avalanche size to the area under a Brownian curve with an
absorbing boundary at the origin,
motivating us to solve this Brownian curve problem.
Thus, we are able to determine the moment generating function for the
avalanche-size probability in this universality class,
explicitly calculating amplitudes of the leading order terms.
\end{abstract}


\maketitle
\section{Introduction}
Sandpile models have played an important role in developing our understanding of self-organized criticality 
\cite{Bak,Dhar1989,Dhar,Priezzhev2001,Mohanty2002}.
One important notion is that of universality, the idea that quantities such as critical
exponents and scaling functions are independent of microscopic details of the model.
This has been studied in the context of individual models, but few have determined
general conditions for models to belong to a particular universality
class \cite{Dhar1989,Mohanty2002,Paczuski1996,Christensen1996}.
In the following, we present details of the solution of a general directed one-dimensional sandpile model introduced in
\cite{Stapleton2005} which is a generalisation of a model studied in \cite{Pruessner2003,Pruessner2004}.
We use a central limit theorem for dependent random variables \cite{Brown1971} to determine
the precise microscopic conditions for scaling of the moments of the avalanche-size probability.
We also argue that there is an $n$-dependent crossover length $\xi_n$, such that
for systems with size $L \ll \xi_n$ branching process behaviour is observed.

The avalanche size statistics are calculated by mapping the model to the problem of finding the
area under a Brownian curve with an absorbing boundary at the origin,
that is, if $x(t)$ is the trajectory of a Brownian curve such that if $x(t') = 0$ for some
$t' > 0$ then $x(t>t') \equiv 0$.
In the large $L$ limit, the avalanche size statistics are identical to those
for the area under the Brownian curve after a ``time'' equal to $L$; $A = \int^L_0 x(t)\rmd t$.
This motivated us to calculate the moment generating function for this area, which is an interesting problem in
its own right as there have been some recent interest in physical applications of the statistics of the area under Brownian curves
\cite{Majumdar, Kearney2005, Kearney2004}.

\section{Definition of The Model}
The model we study is on a one-dimensional lattice of length $L$ where each lattice site, $i = 1,2,\ldots, L$, may be in one of
$n$ states.
The state of site $i$ is denoted $z_i$, which may take values $0,\ldots, n-1$ and this is interpreted as
the number of particles on site $i$.

At the beginning of each time step a particle is added to site $1$: $z_1 \rightarrow z_1 + 1$.
This site may topple a number of times, each toppling redistributing one particle from site $1$ to
site $2$: $z_1 \rightarrow z_1 - 1$, $z_2 \rightarrow z_2 + 1$.
When site $2$ receives a particle it may undergo topplings, redistributing particles to site $3$, which
in turn may topple, and so on until either a site does not topple, or site $L$ topples where the redistributed particles
leave the system and the time step ends.
The avalanche size, $s$ is the total number of topplings which occur during a single time step.
The toppling rules are therefore defined through choosing the probability that a site with $z$ particles will topple
so many times upon receiving a particle.

The only restrictions on the topplings are:
(i) $z_i$ must remain in the range $[0,n-1]$.
For instance, a site with $z_i = 2$ may not topple more that three times when receiving a particle.
Moreover, a site with $z_i = n-1$ which receives a particle must topple at least once..
(ii) When site $i$ topples it redistributes exactly one particle to site $i+1$ only.
(iii) The toppling rule is homogeneous and obeys a Markov property in that the probability that site $i$ topples $s_i$
times depends only on $z_i$.
In fact, the requirement on the homogeneity can be relaxed and all the
following is trivially extended to inhomogeneous toppling rules.
Each site will then have a different stationary state, but provided
that the remaining constraints are obeyed, the scaling of the
avalanches will remain unaltered.
(iv) There must be some probabilistic element to the toppling rules.
To be precise, there must exist at least one value of $z$ such that the number of topplings a site
in this state undergoes is non-deterministic.
This last restriction discounts purely deterministic toppling rules which lead to trivial dynamics.

Self-organised criticality (SOC) is associated with a stationary state where the avalanche-size probability, 
$P(s;L)$, which is the probability of observing an avalanche of size $s$ in a system of size $L$,
obeys simple finite-size scaling,
\begin{equation}
P(s;L) = a s^{-\tau} \mathscr{G}\left( s/bL^{\Delta} \right) \qquad
\text{for } s \gg 1,\ L \gg 1\label{eq:simplescaling}
\end{equation}
where $a$ and $b$ are non-universal constants, $\tau$ and $\Delta$ are universal exponents and $\mathscr{G}$
is a universal scaling function.
It can be shown that if $\lim_{L \rightarrow \infty} \int^{\infty}_{1/bL^{\Delta}} u^{k-\tau} \mathscr{G}(u)\rmd u$ exists and
is non-zero for $k=1,2,\ldots$, then
\begin{eqnarray}
\mom{s}{k}_L &\equiv \sum^{\infty}_{s=1} s^k P(s;L)\nonumber\\
&\propto L^{\Delta(k+1-\tau)}.
\end{eqnarray}
Hence, for $L \gg 1$, the scaling of the moments, $\mom{s}{k}_L$ with $L$ is universal and we can determine the universality
class of a model by calculating the exponents $\Delta$ and $\tau$ from the scaling of these moments.

To formulate the model, we use a Markov matrix representation, which is an extension to the work in \cite{Pruessner2004}.
The configuration of a system is $\{z_1,z_2,\ldots,z_L\}$ for which we shall use the shorthand notation $\{z_i\}$.
In order to construct a Markov matrix representation, we first consider a representation for the configuration of a single site, that
is, a system of size $L=1$.
Such a system can be in one of $n$ stable configurations, $z =
0,1,\ldots,n-1$, and therefore we require an $n$-dimensional vector
space to represent a probability measure over these configurations.
We therefore construct $n$-dimensional left and right basis vectors, $\{\bra{e_z}\}$ and $\{\ket{e_z}\}$ where
\begin{equation}
\bra{e_z} = \left(\rnode{c}{0},\ldots, \rnode{d}{0}, 1, 0, \ldots, 0\right) \quad \text{and} \quad
\ket{e_z} = \left( \begin{array}{c}\rnode{a}{0}\\ \vdots\\ \rnode{b}{0}\\ 1\\ 0\\ \vdots\\ 0\end{array}\right)
\ncbar[nodesep=3pt,angle=0]{a}{b}
\Aput{z \text{ times}}
\ncbar[nodesep=3pt,angle=90]{c}{d}
\Aput{z \text{ times}}
\end{equation}
such that $\scal{e_z}{e_{z'}} = \delta_{z,z'}$.
The configuration of a single site with $z$ particles is then represented by the basis vector $\ket{e_{z}}_1 \equiv \ket{e_{z}}$,
where the subscript $1$ reminds us that it is the configuration of 1 single site.
The $n^L$-dimensional vectors representing configurations $\{z_i\}$ for systems with $L$ sites,
$\ket{e_{\{z_i\}}}_L$, are constructed from these basis vectors
\begin{equation}
\ket{e_{\{z_i\}}}_L = \ket{e_{z_1}}_1 \otimes \ket{e_{z_2}}_1\otimes \ldots \otimes \ket{e_{z_L}}_1
\end{equation}
where $\otimes$ is the usual tensor product.

A state vector $\ket{P_t}_L$ is the weighted sum of basis vectors,
\begin{equation}
\ket{P_t}_L = \sum_{\{z_i\}} w^t_{\{z_i\}} \ket{e_{\{z_i\}}}_L,
\end{equation}
where the weights $w^t_{\{z_i\}}$ are the probabilities that the
system is in the configuration $\{z_i\}$ at time $t$, that is
$w^t_{\{z_i\}} = \scal{e_{\{z_i\}}}{P_t}$.
Note that the normalisation condition requires
\begin{equation}
\sum_{\{z_i\}} w^t_{\{z_i\}} = \sum_{\{z_i\}} \scal{e_{\{z_i\}}}{P_t}_L = 1.
\end{equation}

The system is evolved by applying operators to the state vector $\ket{P_t}_L$.
We define a toppling operator $\op{G}_L$ which adds a particle to site $1$ of a system of size $L$ and
carries out all the topplings:
\begin{equation}
\ket{P_{t+1}}_L = \op{G}_L \ket{P_t}_L,
\end{equation}
making it clear that $\ket{P_t}_L$ is a Markov chain.
Using the $\ket{e_z}$ representation the toppling operator is an $n^L \times n^L$ matrix where 
$\bra{e_{\{z'_i\}}}_L \op{G}_L \ket{e_{\{z_i\}}}_L$ is the
probability that adding a particle to a system in the configuration $\ket{e_{\{z_i\}}}_L$
results in the configuration $\ket{e_{\{z'_i\}}}_L$ after topplings.
See Appendix \ref{sec:precise} for an explicit representation of $\op{G}_L$ for a system with $n=2$.
The steady state, $\ket{0}_L$ is defined as the state which is
invariant under application of the toppling operator $\op{G}_L$.
It is therefore the right eigenvector of $\op{G}_L$ with eigenvalue $1$,
\begin{equation}
\op{G}_L \ket{0}_L = \ket{0}_L.
\end{equation}
The corresponding left eigenvector, $\bra{0}_L$ satisfies
\begin{equation}
\bra{0}_L \op{G}_L = \bra{0}_L.
\end{equation}
Since particle number is conserved in the bulk, the sum of each column in $\op{G}_L$ equals 1.
This leads to
\begin{equation}
(1,1,\ldots,1) \op{G}_L = (1,1,\ldots,1),
\end{equation}
identifying the left eigenvector of the toppling operator $\bra{0}_L = (1,1,\ldots,1)$.

In order to calculate the moment generating function for the avalanche-size probability
it is convenient to define the operator $\op{G}_L(x)$ such that
\begin{equation}
Q_{L,m}(x) = \bra{0}_L \left[ \op{G}_L(x)\right]^m\ket{0}_L
\end{equation}
is the avalanche size moment generating function over $m$ time steps and
\begin{equation}
\langle s^k \rangle_L = \left. \left( x \frac{\rmd}{\rmd x}\right)^k Q_{L,1}(x) \right|_{x=1} \equiv Q^{(k)}_{L,1}
\end{equation}
gives the $k$th moment of the avalanche-size probability in a system of size $L$.
The toppling operator is then
\begin{equation}
\op{G}_L \equiv \op{G}_L(1).
\end{equation}
To illustrate how to construct $\op{G}_L(x)$, consider the action of adding a particle to site $i=1$.
First, we look at the number of particles on that site, $z_1$, and determine the number of times the site topples
using the probabilities given by the toppling rule.
Hence, we require $n+1$ matrices of dimension $n$, denoted $\op{S}_k$, which will act on the first site and remove 
$k$ particles, multiplying the final state by the probability that this toppling took place.
We then need to redistribute these particles to site $i=2$, which we achieve by acting $(\op{G}_{L-1})^k$ on the remaining
$L-1$ sites.
This works because adding to site $2$
of a system of size $L$ is equivalent to adding to site $1$ of a system of size $L-1$ (note $\op{G}_0(x) \equiv \Id$).
Finally, we multiply the remaining state by a factor $x^k$, which marks the state as having toppled $k$ times, which gives
moments of the avalanche size upon differentiation.
This leads us to write the general toppling operator
\begin{equation}
\op{G}_L(x) = \sum^n_{k=0} x^k \left[ \Id \otimes \op{G}_{L-1}(x)\right]^k\op{S}_k \otimes \Id^{\otimes L-1}\label{eq:genop},
\end{equation}
where $\Id$ is an $n\times n$ identity matrix and $\op{A}^{\otimes N} \equiv \op{A} \otimes \op{A} \otimes \ldots \otimes \op{A}$, $N$ times.
The restrictions, (i)-(iv), on the model give $[\op{S}_k]_{ij} \geq 0$ for $j = i + k -1$ and
equal to zero otherwise, that is,
\begin{equation}
\op{S}_k = \sum_z \ket{e_{z+k-1}}_1 S_{z,z+k-1} \bra{e_z}_1
\end{equation}
where the sum is over all $z$ which satisfy both $0 \leq z \leq n-1$ and $0 \leq z + k - 1 \leq n-1$ and
$S_{z,z+k-1}$ is the probability that a site with $z$ particles topples $k$ times on receiving a particle.
Note that particle conservation requires $\sum^{n-1}_{j=0} S_{z,j} = 1$

\section{Stationary Properties}
\label{sec:stationary}
In this section we find the steady state, $\ket{0}_L$, which is the eigenvector of $\op{G}_L$ with eigenvalue 1.
Consider the single site operator,
\begin{equation}
\op{G}_1(x) = \sum^n_{k=0} x^k \op{S}_k.
\end{equation}
We begin by finding the eigenvectors and eigenvalues defined by
\numparts
\begin{eqnarray}
\bra{\lambda_i(x)} \op{G}_1(x) = \lambda_i \bra{\lambda_i(x)} \\
\op{G}_1(x)\ket{\lambda_i(x)} = \lambda_i \ket{\lambda_i(x)}
\end{eqnarray}
\endnumparts
where $i$ takes values from $0$ to $n-1$.
From the properties of the $\op{S}_k$ and the normalisation condition we find
\begin{equation}
\bra{\lambda_0(x)} = \left(\frac{1}{x^{n-1}}, \frac{1}{x^{n-2}}, \ldots, \frac{1}{x}, 1\right)
\end{equation}
satisfies
\begin{equation}
\bra{\lambda_0(x)} \sum^n_{k=0} x^k \op{S}_k = x \bra{\lambda_0(x)}\label{eq:lefteig}
\end{equation}
and so is the left eigenvector of $\op{G}_1(x)$ with eigenvalue $\lambda_0 = x$.
The corresponding right eigenvector therefore satisfies
\begin{equation}
\op{G}_1(x) \ket{\lambda_0(x)} = \sum^n_{k=0} x^k \op{S}_k \ket{\lambda_0(x)} = x \ket{\lambda_0(x)},\label{eq:righteig}
\end{equation}
which determines the precise form of the eigenvector.
As the eigenvectors must be normalised, $\scal{\lambda_0(x)}{\lambda_0(x)} = 1$,
we may write
\begin{equation}
\ket{\lambda_0(x)} = 
\left(\begin{array}{c}
p_0 x^{n-1}\\
p_1 x^{n-2}\\
\vdots\\
p_{n-1} x^0
\end{array}
\right)
\end{equation}
where $p_z$ is the probability that a site contains $z$ particles in the stationary state and $\sum^{n-1}_{z=0} p_z = 1$.
We cannot, however, determine $p_i$ any more precisely without details of the $\op{S}_k$ and these
will have to be calculated separately in each case.

If the matrix $\op{G}_1$ is a regular Markov matrix, that is, there exists an integer $N \geq 1$ such that $[\op{G}_1^N]_{ij} > 0$ for all $i,j$,
then we have found the unique stationary state of the single site operator,
\begin{equation}
\ket{0} = \ket{\lambda_0(1)} =
\left(\begin{array}{c}
p_0 \\
p_1 \\
\vdots\\
p_{n-1}
\end{array}
\right).
\end{equation}
In the following we shall always assume that $\op{G}_1$ is regular.
The discussion of the necessary and sufficient conditions for a regular $\op{G}_1$ is non-trivial
and we shall not discuss it in detail.
We shall simply note that this requirement, along with restrictions (i)-(iv), still leaves an abundance
of choice for the toppling rules.
For instance, it is easy to demonstrate that any tridiagonal matrix with positive definite elements is regular.
Hence, any toppling rule which always allows a site to topple zero times, once or twice on receiving
a particle (with the usual exceptions for $z=0$ and $z=n-1$) will automatically lead to an acceptable toppling rule.

Now that we have found the stationary state of the single site operator $\op{G}_1$, we shall proceed to determine the stationary properties of
the full operator $\op{G}_L$ by induction.
We introduce the notation $\ket{\lambda_i(x)}_L$ as the $i$th eigenvector of $\op{G}_L(x)$ such that
\begin{equation}
\op{G}_L(x)\ket{\lambda_i(x)}_L = \lambda_{L,i}(x)\ket{\lambda_i}_L,
\end{equation}
where $\lambda_{L,i}(x)$ is the $i$th eigenvalue of $\op{G}_L(x)$.
We now make the ansatz that these eigenvectors may be expressed
\begin{equation}
\ket{\lambda_{i+j}(x)}_L = \ket{\lambda_j\bigl(x\lambda_{L-1,i}(x)\bigr)}_1\otimes \ket{\lambda_i(x)}_{L-1}.
\end{equation}
Operating on the left hand side with $\op{G}_L(x)$ we find
\begin{eqnarray}
\op{G}_L(x) \ket{\lambda_{i+j}(x)}_L &= \sum_{k=0}^n (x \lambda_{L-1,i})^k\op{S}_k \ket{\lambda_{i+j}(x)}_L\\
&= \op{G}_1(x\lambda_{L-1,i}) \ket{\lambda_j\bigl(x\lambda_{L-1,i}(x)\bigr)}_1\otimes \ket{\lambda_i(x)}_{L-1}\\
&= x \lambda_j(x\lambda_{L-1,i})\ket{\lambda_{i+j}(x)}_L
\end{eqnarray}
So, we find that $\ket{\lambda_{i+j}(x)}_L$ is indeed an eigenvector of $\op{G}_L(x)$ with eigenvalue
$\lambda_{L,i+j} = \lambda_j(x\lambda_{L-1,i})$.
Hence, by induction, and recalling that we have assumed $\op{G}_L$ is regular, the unique steady state is
\begin{equation}
\ket{0}_L = \ket{0}_1 \otimes \ket{0}_{L-1} = \ket{0}^{\otimes L}_1.
\end{equation}
This is a product state, which means that it has no spatial
correlations and indeed we find that avalanches in small systems are
uncorrelated, leading to branching process behaviour.
However, we will show that in larger systems temporal correlations
develop which bring the avalanche behaviour away from the branching
process to that characterised by the area under a Brownian curve.

\section{Toppling probability distribution}
In this section we define the toppling probability distribution and determine some of its properties which
will be used in the next section.
The toppling probability distribution, $P(s;L,m)$ is defined as the probability that a system of size $L$ in the stationary
state undergoes a total of $s$ topplings on receiving $m$ particles.
In principle, it may be calculated from the moment generating function
\begin{equation}
P(s;L,m) = \frac{1}{s!} \left. \frac{\rmd^{s}}{\rmd x^{s}} \bra{0}_L \left[ \op{G}_L(x)\right]^{m} \ket{0}_L\right|_{x=0}
\end{equation}
although this is rarely a simple task in practice.
Recall that we only consider toppling rules obeying the restrictions (i)-(iv) with a unique
stationary state.  First, we show that $P(s;L,m)$ has a mean value
\begin{equation}
Q^{(1)}_{L,m} = \sum^{\infty}_{s=0}\ s\ P(s;L,m) = mL,
\end{equation}
which is what we would expect by considering conservation in the stationary state since
every particle that enters the system must leave through the open boundary.
We write down the equation for the first moment
\begin{eqnarray}
Q^{(1)}_{L,m} &= \left. \frac{\rmd}{\rmd x} \bra{0}_L \left[\op{G}_L(x)\right]^m \ket{0}_L \right|_{x=1} \nonumber\\
&= m \sum^n_{k=0} \bra{0}_1 k \op{S}_k \ket{0}_1 \left( 1 + Q^{(1)}_{L-1,1}\right).
\end{eqnarray}
Multiplying \eref{eq:righteig} on the left by $\bra{\lambda_0(x)}$ and differentiating, we find
\begin{equation}
\bra{\lambda_0(x)} \sum^n_{k=0} k x^{k-1} \op{S}_k \ket{\lambda_0(x)} = 1
\end{equation}
and so $\bra{0}_1 \sum^n_{k=0} k \op{S}_k \ket{0}_1 = 1$
giving $Q^{(1)}_{L,m} = m ( 1 + Q^{(1)}_{L-1,1})$, with $Q^{(1)}_{1,1} = 1$, which has the solution $Q^{(1)}_{L,m} = mL$.

Next, we show that the avalanche probability may be factorized.
We define $P(s,t; m, 1, L)$ as the joint probability that a system of size $1+L$, which has received $m$ particles, undergoes
$s$ topplings in the first site and $t$ in the remaining $L$ sites.
\begin{equation}
P(s,t;m,1,L) = \frac{1}{s!} \left. \frac{\rmd^{s}}{\rmd x_1^{s}}
\frac{1}{t!} \frac{\rmd^{t}}{\rmd x^{t}_L}
\bra{0}_{L+1} \left[ \op{G}_{L+1}(x_1,x_L)\right]^m \ket{0}_{L+1}\label{eq:joint}\right|_{x_1=0\\x_L=0}
\end{equation}
where
\begin{equation}
\op{G}_{L+1}(x_1,x_L) \equiv \sum^n_{k=0} x_1^k \op{S}_k \otimes \left[ \op{G}_L(x_L) \right]^k.
\end{equation}
By expanding the bracket in \eref{eq:joint} and carrying out the differentiation with respect to $x_1$, we find
\begin{eqnarray}
\fl P(s,t;m,1,L) = \bra{0}_1 \sum_{\{k_i\}} \delta\left(\sum_i k_i - s\right) \prod^m_{i=0} \op{S}_{k_i} \ket{0}_1\nonumber \\
\quad \frac{1}{t!} \left. \frac{\rmd^{t}}{\rmd x_L^{t}} \right|_{x=0} \bra{0}_L \left( \op{G}_L(x_L)\right)^{s} \ket{0}_L
\end{eqnarray}
where $\delta(x)$ is the Kronecker delta, $\delta(0) = 1$ and $\delta(x) = 0$ for $x \neq 0$.
Identifying the first scalar product as simply the probability that a site charged $m$ times topples $s$ times,
$P(s;m,1)$, we have
\begin{equation}
P(s,t;m,1,L) = P(s;m,1)P(t;s,L).\label{eq:factor}
\end{equation}
This is simply a statement about the fact that the directed nature means that sites $i=2\ldots L$ do not have any
influence on site $i=1$.
Hence, if we consider $s_i$, which is the number of times site $i$ topples during a particular avalanche, this result
tells us that in the stationary state the sequence $s_1,s_2,s_3,\ldots,s_L$ for a single avalanche forms a Markov chain.
This will become important later on when we come to map the avalanche
size $s = \sum^L_{i=1} s_i$ to the area under a random walker.

We now derive three important results for the single site toppling probability distribution, $P(s;1,m)$,
which we will use later to make the mapping of avalanches to the area under a Brownian curve more rigorous.
First, we show that the range of $s$ for which $P(s;1,m)$ has support has an upper bound equal to $2n-1$.
Second, we show that $P(s;1,m)$ has a stationary distribution for large $m$,
\begin{equation}
\lim_{m\rightarrow \infty} P(s;1,m) = \sum_z p_z p_{(m-s)+z}
\end{equation}
where the sum is over all $0 \leq z \leq n-1$ satisfying $0 \leq m - s + z \leq n-1$.
Note that this is a function of $m-s$ only.
Finally, we consider the width of $P(s;1,m)$ around its mean value,
\begin{equation}
\tilde{Q}^{(2)}_{1,m} \equiv \sum^{\infty}_{s=0} (s-m)^2 P(s;1,m)\label{eq:Qtilde}
\end{equation}
and show that $\tilde{Q}^{(2)}_{1,m} > 0$ for all $m$.
These results lead to a further result that the width 
is finite and non-zero for all $m$, and approaches a constant for $m \rightarrow \infty$, 
which, again, will become important later when we map the problem to a random walker.

The first of these results follows immediately from bulk conservation of particles:
Consider a site with $z$ particles, which is charged $m$ times.
After $s$ topplings have taken place it will have $z' = z + m - s$ particles.
Since both $z$ and $z'$ must lie between $0$ and $n-1$, $P(s;1,m)$ may only have
non-zero values for $m-n+1 \leq s \leq m + n - 1$.
Hence, $P(s;1,m) = 0$ for $|s-m| > n-1$ since such topplings are always impossible and
$\tilde{Q}^{(2)}_{1,m} \leq (n-1)^2 < \infty$ for $n < \infty$.
Hence, $\tilde{Q}^{(2)}_{1,m}$ is finite because the sum in \eref{eq:Qtilde} has at most $2n-1$
non-zero terms.

Next, we consider the probability distribution, $P_m(z'|z)$,
which is the probability that a site having $z$ particles which is charged $m$ times, is left with
$z'$ particles after $s = z' - z + m$ topplings.
In the stationary state, this is related to $P(s;1,m)$ by
\begin{equation}
P(s;1,m) = \sum^{n-1}_{z=0} p_z P_m(m+z-s|z).
\end{equation}
Since the values of $z$ a site passes through as it topples is a Markov chain, 
we may write down the Chapman-Kolmogorov equation for $P_m(z'|z)$,
\begin{equation}
P_{m+1}(z'|z) = \sum^{n-1}_{z''=0} P_1(z'|z'') P_m(z''|z).\label{eq:Ch-K}
\end{equation}
The probabilities $P_1(z'|z'')$ are related to the single site operator, $\op{G}_1$ by
$P_1(z'|z'') = \bra{e_{z'}}_1 \op{G}_1 \ket{e_{z''}}_1$ and so
we introduce the matrix $\op{P}_m$, such that
$P_m(z'|z'') = \bra{e_{z'}}_1 \op{P}_m \ket{e_{z''}}_1$
and \eref{eq:Ch-K} becomes
\begin{eqnarray}
\bra{e_{z'}}_1\op{P}_{m+1}\ket{e_{z}}_1 &= 
\sum^{n-1}_{z''=0} \bra{e_{z'}}_1 \op{G}_1 \ket{e_{z''}}_1 \bra{e_{z''}}_1 \op{P}_m \ket{e_{z}}_1\nonumber\\
&= \bra{e_{z'}}_1\op{G}_1\op{P}_m\ket{e_{z}}_1.
\end{eqnarray}
Thus $\op{P}_{m+1} = \op{G}_1 \op{P}_m$ and, since $\op{G}_1$ is a regular Markov matrix,
there will be a unique stationary distribution $\op{P}_{\infty}$ satisfying
\begin{equation}
\op{P}_{\infty} = \op{G}_1 \op{P}_{\infty}.\label{eq:matPinf}
\end{equation}
The only non-trivial solution to \eref{eq:matPinf} is
\begin{equation}
\op{P}_{\infty} = \left( \ket{0}_1, \ket{0}_1, \ldots, \ket{0}_1\right)
\end{equation}
where we have noted that the stationary state, $\ket{0}_1$, is unique and conservation of probability
requires
$\sum^{n-1}_{z'=0} \bra{e_{z'}}_1 \op{P}_{\infty} \ket{e_{z}}_1 = 1$
This leads us to
\numparts
\begin{eqnarray}
\lim_{m \rightarrow \infty} P_m(z'|z) &= p_{z'}\\
\lim_{m \rightarrow \infty} P(s;1,m) &= \sum_z\ p_z\ p_{(m-s)+z}\label{eq:Pinf}
\end{eqnarray}
\endnumparts
where we have noted that $\scal{e_{z'}}{0}_1 = p_{z'}$ and the sum is over $0 \leq z \leq n-1$ satisfying $0 \leq m-s+z \leq n-1$.
Finally, \eref{eq:Pinf} leads to
\begin{equation}
\lim_{m\rightarrow \infty}\tilde{Q}^{(2)}_{1,m} = \sum^{n-1}_{s=-n+1} s^2 \sum_z\ p_z\ p_{z-s}
\end{equation}
where the sum is over all $0 \leq z \leq n-1$ such that $0 \leq z+s \leq n-1$.
This limit exists, is non-zero and is finite for $n < \infty$.

Finally, we show that $\tilde{Q}^{(2)}_{1,m} > 0$ for all $m$.
Consider again the probability distribution
$\op{P}_m = \left[\op{G}_1\right]^m$
and the equation for the width,
\begin{equation}
\!\tilde{Q}^{(2)}_{1,m}\! =\! \sum^{\infty}_{s=0} (s-m)^2 \sum^{n-1}_{z=0}\ p_z\ P_m(m+z-s|z)\label{eq:widthpz}
\end{equation}
For $\tilde{Q}^{(2)}_{1,m^*} = 0$, for some $m^*$, we require
\begin{equation}
p_z \bra{e_{z'}}_1 \op{P}_{m^*} \ket{e_{z}}_1 = p_z \delta_{z,z'}
\end{equation}
which follows from normalisation of $\op{P}_m$ and the fact that \eref{eq:widthpz} has no negative terms.
This implies that, for all $0 \leq z \leq n-1$ for which $p_z > 0$,
\begin{equation}
\left[ \op{G}_1 \right]^{N m^*} \ket{e_z}_1 = \ket{e_z}_1\label{eq:cont}
\end{equation}
where $N>0$ is an integer.
However, since $\op{G}_1$ is regular, there exists an integer $N^*$ such that there is only one vector, $\ket{0}_1$, 
satisfying
\begin{equation}
\left[ \op{G}_1 \right]^{N} \ket{0}_1 = \ket{0}_1
\end{equation}
for any $N > 0$.
If there are more than one values of $z$ for which $p_z > 0$ this contradicts \eref{eq:cont}, and so $\tilde{Q}^{(2)}_{1,m}$ is never zero.
If, however, we have a single value, $z^*$ such that $p_z = \delta_{z,z^*}$, then $\ket{0}_1 = \ket{e_{z^*}}$ and \eref{eq:cont} does not
lead to a contradiction.
However, in this case the dynamics are trivial as the steady state has all sites with exactly $z^*$ particles and any particle
added to the system will pass through immediately with exactly $L$ topplings.

\section{Mapping To Area Under Random Walker}
We now come to the main result we need in order to determine the avalanche statistics for the directed sandpile,
which is that it may be mapped exactly onto a random walker on $[0,\infty)$ with an absorbing
boundary at the origin.
After adding a particle at the beginning of a time step, site $i=1$ will topple $s_1 \geq 0$ 
times with probability $P(s_1;1,1)$.
These $s_1$ particles are redistributed to site $i=2$, which will topple $s_2 \geq 0$ times with probability
$P(s_2;1,s_1)$.
The probability of site $2$ toppling $s_2$ times, independent of $s_1$, which we denote $\phi_2(s_2)$, is
\begin{equation}
\phi_2(s_2) = \sum^{\infty}_{s_1 = 1} P(s_2;1,s_1)
\end{equation}
which follows from \eref{eq:factor}.
Defining $\phi_i(x)$ as the probability that site $i$ topples $x$ times independent of previous topplings, we have
\begin{equation}
\phi_{i+1}(x) = \sum^{\infty}_{y = 1} \phi_i(y)P(x;1,y) \qquad
\text{for } i=1,\ldots,L-1.\label{eq:directed_walker}
\end{equation}
This is a random walker on the interval $[0,\infty)$ with the probability of hopping from $y$ to $x$ equal to $P(x;1,y)$.
There is an absorbing boundary at $x=0$ since any non-toppling site stops the avalanche.
If we denote the trajectory $x(i)$, $i=0\ldots L$, then the avalanche size is
\begin{equation}
s = \sum^L_{i=1} x(i)
\end{equation}
with $x(0) = 1$, which is the area under the trajectory $x(i)$.

Note that the random walker described by \eref{eq:directed_walker} has jumps which are correlated since
the probability of hopping from $y$ to $x$ depends explicitly on $y$ and $x$, and not simply the difference $x-y$.
This means we must be careful if we wish to use the results for the uncorrelated random walker, or its continuum limit.
However, in Ref.~\cite{Brown1971}, the author remarks that for martingales with a fixed maximum jump size
exhibiting stationarity and ergodicity, there is a quantity, $s_i^2 = \mathds{E}\sum^i_{n=1}{\sigma_n}^2$ such that
\begin{equation}
\!\lim_{i\rightarrow \infty}\! P[x(i)/s_i \leq x]\! =\! (2\pi)^{-1}\! \int^x_{-\infty}\!\! e^{-\frac{1}{2}y^2}\ \rmd y\label{eq:normal}
\end{equation}
where ${\sigma_n}^2$ is the variance of the $n$th step in the process.

To apply this result, we extend the random walker described by $P(x;1,y)$ to the full space, $(-\infty,\infty)$ for
we may add in the effect of the boundaries later by use of mirror charges \cite{Rudnick2004}.
As we have assumed the existence of a unique stationary state and have proven that $Q^{(1)}_{1,m} = m$ and
$0 < \tilde{Q}^{(2)}_{1,m} \leq (n-1)^2$, all that is left to prove is ergodicity.
This is equivalent to showing that the set of recurrent
states of the random walker are irreducible, that is, the probability of reaching
any recurrent state $i$ from any other recurrent state $j$ is non-zero.
Two states, $i$ and $j$ which have this property are said to intercommunicate, denoted $i \leftrightarrow j$.
We consider the fact that $\op{G}_1$ is assumed to be regular, in which case
there exists an $N$ such that $\bra{e_z} [\op{G}_1]^m \ket{e_{z'}} > 0$ for all $z,z'\in [0,n-1]$ and $m\geq N$.
Hence, all states $i,j \geq N$ intercommunicate since $P(m\pm 1;1,m) > 0$ for all $m \geq N$.
We also note that $0\leftrightarrow N$ and $1 \leftrightarrow N$ which follow respectively because the avalanche should
always be able to finish in an infinite system and arbitrarily large avalanches can be initiated from a single
added particle.
When we consider states $k<N$, we note that there can only be a finite number of these which do not intercommunicate with
state 1.
Since there is a unique stationary state which includes all states $i \geq N$, these non-intercommunicating states must
be transient and ergodicity follows.
Hence, we have now proven that for a toppling rule obeying the restrictions (i)-(iv) with a unique stationary state,
(i.e $\op{G}_1$ is regular), the distribution of the random walker on $(-\infty,\infty)$ will approach the normal distribution
given by \eref{eq:normal}.
This means that, for long times, such a random walker with dependent jump sizes will have the statistics
of ordinary diffusion with diffusion constant $2D = s_n$.
Hence, by adding mirror charges to remove paths that cross $x=0$, we are able to calculate the large $L$ statistics
of avalanches directly from the area under the Brownian curve,
which is our justification for calculating moments in the continuum limit in the next section.
Of course, we could have simply gone ahead and carried out the calculations in the continuum without the above
analysis and demonstrated that they correctly modelled the numerics.
However, had we done so we would not have had a precise idea of how trustworthy these calculations were and where we
expect them to break down.

\section{Moments of the area under the Brownian curve}
\label{sec:moments}
Having proven the correspondence between avalanches and a random walk of independent identically distributed step sizes, we proceed
to calculate the moment generating function for the area under the Brownian curve.
The authors are aware of only one study which investigates the finite-size effects due to stopping the
curve after some time, which corresponds to the finite size of the sandpile \cite{Pruessner2004} and
since our analysis goes further than that in Ref.~\cite{Pruessner2004}, we present it here in some detail.
The following calculation will be carried out using notation and language suitable for the random walker
description of the problem.
Hence, the Brownian curve will be described by a trajectory $x(t)$ where $x$ is interpreted as ``space'' and $t$ is
``time'' with the diffusion constant $D$ having units $\unit{\text{Length}}^2\unit{\text{Time}}^{-1}$.
We do this because the path integral approach we are about to employ is more intuitive in this language.\footnote{
We should stress that in this picture we consider the Brownian curve $x(t)$ as existing on the entire interval
$[0,\infty)$ and we measure the area up to the point $L$, $A = \int^L_0 x(t) \rmd t$.
Hence, what is a boundary in the sandpile picture (the open boundary at site $i=L$) is not considered
a boundary in the Brownian curve picture.}
The connection to the sandpile is made by noting that the number of topplings of site $i$ is equal to
$x(t=i)$ and the system size, $L$, is equal to time at which we stop the curve $x(t)$.

\begin{figure}
\centering
\epsfig{file=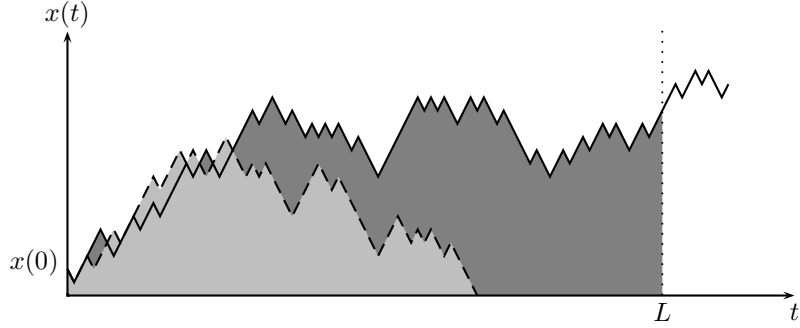,clip=}
\label{fig:brownian}
\caption{The area under the Brownian curve with an absorbing boundary.
The statistics consist of contributions due to curves $x(t)$ which are nonzero at $t=L$ (solid line), as well as those which cross the boundary
at some time $t < L$ (dashed line).}
\end{figure}
We begin with the generating function
\begin{equation}
\langle e^{-\lambda A}\rangle = \int^{\infty}_0 e^{-\lambda A} P(A;x,L)\rmd A
\end{equation}
where $P(A;x,L)$ is the probability that a random walker starting at $x$ has the area under its trajectory
equal to $A$ after time $L$.
If we denote the trajectory of the walker $x(t)$, then the curves contributing to $P(A;x,L)$ are all those
which satisfy
\begin{equation}
\int^L_0 x(t) \rmd t = A.
\end{equation}
Note that we have an absorbing boundary at $x=0$, such that if $x(t') = 0$ for any $t$ then $x(t>t') \equiv 0$.
Hence, there are two contributions to $P(A;x,L)$: That due to trajectories which do not cross the absorbing boundary,
$x(L) > 0$ and those which cross the absorbing boundary at some time $t \leq L$, see figure~\ref{fig:brownian}.

We shall treat these separately, writing
\begin{eqnarray}
\av{e^{-\lambda A}} &= \int^{\infty}_0 \rmd A \int^{\infty}_0 \rmd y\ e^{-\lambda A} \Psi(A,y,L;x)
+ \int^{\infty}_0 \rmd A \int^{L}_0 \rmd t\ e^{-\lambda A} \Phi(A,t;x)\label{eq:genpart}\nonumber\\
&\equiv I_1 + I_2
\end{eqnarray}
where $\Psi(A,y,L;x)$ is the probability that a trajectory beginning at $x(0) = x$ passes through
$x(L) = y$ with area $A$ and $\Phi(A,t;x)$ is the probability that a trajectory beginning at
$x(0) = x$ first touches the absorbing boundary at time $t$, with area $A$.

Using standard path integral methods, we may write down
\begin{equation}
\fl \Psi(A,y,T;x) = \lim_{g \rightarrow \infty} \mathop{\int_{x(0) = x}}_{x(L) = y}
 \mathscr{D}x(t) \delta \left(\int^L_0 x(t)\rmd t - A\right)
\exp\left( -\int^T_0 [D\dot{x}^2 + g\delta(x)] \rmd t\right)
\end{equation}
where $\dot{x} = \rmd x(t)/\rmd t$.
Taking the integral over $A$ we find that the first term on the right hand side of \eref{eq:genpart} is
\begin{equation}
I_1 = \lim_{g \rightarrow \infty}\int^{\infty}_0 \rmd y\ \mathop{\int_{x(0) = x}}_{x(L) = y} \mathscr{D}x(t)
\exp\left( -\int^L_0 \left[ D\dot{x}^2 + \lambda x + g\delta(x)\right] \rmd t\right).
\end{equation}
Following the lines of Ref.~\cite{Majumdar}, we note that this is simply the path integral for a Brownian particle
with a linear potential for $x\in (0,\infty)$ and an infinite potential at $x=0$.
Hence, we write this term as
\begin{equation}
I_1 = \lim_{g \rightarrow \infty}\int^{\infty}_0 \bra{y} e^{-\hat{H}L} \ket{x}\rmd y
\end{equation}
where $\hat{H} = -D\fracpartn{x}{2} + \lambda x + g\delta(x)$.
The resulting equation of motion, $\fracpart{t} \ket{\phi} = - \hat{H} \ket{\phi}$ is
easily solved using Airy functions which can be used to form an orthonormal basis on $[0,\infty)$ \cite{Vallee2004},
\begin{equation}
I_1 = \sum^{\infty}_{j=1} \frac{\Ai\left( \left(\frac{\lambda}{D}\right)^{1/3}x + x_j\right) \int^{\infty}_{x_j} \Ai(z)\rmd z}{\Ai'(x_j)^2}
e^{x_j \lambda^{2/3}D^{1/3}L}
\end{equation}
where $x_j$ are the zeros of the Airy function, $x_1 = -2.338\ldots$, $x_2 = -4.087\ldots$ etc.

In a similar way, for the second term on the right hand side of \eref{eq:genpart}, we have
\begin{equation}
\Phi(A,t;x) = D\fracpartat{y}{0}{\Psi(A,y;x,t)}
\end{equation}
since this is the current of diffusing particles with area under the curve equal to A, leaving the system at time $t$.
Hence
\begin{eqnarray}
I_2 &= \int^{L}_0 \fracpartat{y}{0}{\bra{y} e^{-\hat{H}t} \ket{x}}\rmd t\nonumber\\
&= \int^L_0 \lambda^{2/3} D^{1/3}
\sum^{\infty}_{j=1} \frac{\Ai\left( \left(\frac{\lambda}{D}\right)^{1/3}x + x_j\right)}{\Ai'(x_j)}
e^{x_j \lambda^{2/3}D^{1/3}t}\rmd t.
\end{eqnarray}

In order to proceed, we use the fact that the leading order $L$ dependence for each moment
come from terms linear in $x$.
In Ref.~\cite{Pruessner2004} it was shown that if the moment generating function is written,
\begin{eqnarray}
\langle A^n\rangle &= (-1)^n \left. \frac{\partial^n}{\partial \lambda^n} \av{e^{-\lambda A}} \right|_{\lambda = 0}\nonumber\\
&\equiv n!\ \phi_n(x,L),
\end{eqnarray}
then $\phi_n(x,L)$ may be determined recursively
\begin{equation}
\phi_n(x,L) = \int^{\infty}_0 \rmd x' \int^L_0 \rmd t\ x' \phi_{n-1}(x',t) G(x,L-t;x'),
\end{equation}
where $G(x,t;x')$ is the propagator for the diffusion equation with appropriate boundaries,
\begin{equation}
G(x,t;x') = \frac{e^{-\frac{(x-x')}{4Dt}^2} - e^{-\frac{(x+x')}{4Dt}^2}}{\sqrt{4\pi D t}}.
\end{equation}
If we define the ``current''
\begin{equation}
j_n(L) = \fracpartat{x}{0}{\phi_n(x,L)}
\end{equation}
then if $j_n(L)$ is non-zero, $\phi_n(x,L)$ is proportional to $x$ to lowest order.
In this case
\begin{equation}
j_n(L) =\int^{\infty}_0 \rmd x' \int^{L}_0 \rmd t \frac{x'^2}{\sqrt{4\pi}} \frac{\phi_{n-1}(x',t)}{[D(L-t)]^{3/2}} e^{-\frac{x'^2}{4D(L-t)}}
\end{equation}B
and so $j_n(L) > 0$ for $n > 0$ since the integrand is always positive definite.
Hence, all moments are proportional to $x$ to lowest order.
The fact that the terms linear in $x$ will also be the highest order in $L$ follows from dimensional analysis.
If we write down the expansion of a moment in powers of $x$ then each term must have the same dimension.
By considering the dimensions of the available quantities, such an expansion must take the form
\begin{equation}
\av{A^n} = x C_n D^{(n-1)/2} L^{(3n-1)/2} + \sum^{\infty}_{k=2} x^k C_{n,k} D^{(n-1-2k)/2} L^{(3n-1-2k)/2}
\end{equation}
where $C_{n,k}$ are simply more coefficients with no $x$, $D$ or $L$ dependence.
Hence, the term of lowest order in $x$ will have the highest order $L$ dependence.

Taylor expanding $I_1$ and $I_2$ to first order about $x=0$,
\numparts
\begin{eqnarray}
I_1 &\approx x \left( \frac{\lambda}{D}\right)^{1/3}
\sum^{\infty}_{j=1} \frac{\int^{\infty}_{x_j} \rmd z\ \Ai(z)}{\Ai'(x_j)}e^{x_j \lambda^{2/3}D^{1/3}L} \equiv J_1\label{eq:J1}\\
I_2 &\approx x \lambda \int^L_0 \sum^{\infty}_{j=0} e^{x_j \lambda^{2/3} D^{1/3} t} \rmd t\equiv J_2\label{eq:J2}.
\end{eqnarray}
\endnumparts
Note, however, that this approximation is not valid for the zeroth moment, $\av{A^0} = 1$ since it is not proportional to $x$.
$J_1$ and $J_2$ are now in similar forms to equations appearing in Ref.~\cite{Majumdar}.
They calculate the quantity
\begin{eqnarray}
\tilde{P}(\lambda,L) &= \sqrt{\pi}2^{-1/6}(\lambda L^{3/2})^{1/3}
\sum^{\infty}_{j=1} \frac{\int^{\infty}_{x_j} \Ai(z)\rmd z}{\Ai'(x_j)}e^{x_j \lambda^{2/3}2^{-1/3}L}\label{eq:Majumdar1}\nonumber\\
&= \sum^{\infty}_{n=0} \frac{(-\lambda)^n}{n!} a_n L^{3n/2}
\end{eqnarray}
where the $a_n$ have been calculated in Ref.~\cite{Perman1996}.
We simply quote the first few values,
\begin{equation}
a_0 = 1, \quad a_1 = \frac{3}{4}\sqrt{\frac{\pi}{2}}, \quad a_2 = \frac{59}{60}, \quad a_3 = \frac{465}{512}\sqrt{\frac{\pi}{2}},
\quad a_4 = \frac{5345}{3696}.
\end{equation}
Apart from a few multiplicative prefactors, \eref{eq:Majumdar1} differs from \eref{eq:J1} only by the fact that the
former uses $D = 1/2$.
We therefore have to reinsert the diffusion constant, $D$, which they assumed equal to $1/2$, but this is easily done
by considering the dimensions of the results.
We note that $J_1$ is a dimensionless function, and so
\begin{equation}
J_1 = \left(\frac{\frac{1}{2}}{D}\right)^{\gamma} x \left(\frac{\lambda}{D}\right)^{1/3} 
\frac{2^{1/6}}{\sqrt{\pi}} (\lambda L^{3/2})^{-1/3} \tilde{P}(\lambda,L)
\end{equation}
where $\gamma$ is chosen such that $J_1$ is dimensionless.
It is then easy to show that $\gamma = 1/6-n/2$ and
\begin{equation}
J_1 = x \sum^{\infty}_0 \frac{(-\lambda)^n}{n!} c_n D^{(n-1)/2} L^{(3n-1)/2}
\end{equation}
where
\begin{equation}
c_n = \frac{2^{n/2}a_n}{\sqrt{\pi}}.
\end{equation}

We may carry out an identical procedure for $J_2$.
The equivalent quantity in Ref.~\cite{Majumdar} is
\begin{eqnarray}
\tilde{P}(\lambda,L) &= \sqrt{2\pi} (\lambda L^{3/2}) \sum^{\infty}_{j=0} e^{x_j \lambda^{2/3} 2^{-1/3} L}\nonumber\\
&= \sum^{\infty}_{n=0} \frac{(-\lambda)^n}{n!} b_n L^{3n/2}
\end{eqnarray}
where, again, $b_n$ have been calculated in Ref.~\cite{Perman1996}, the first few values being
\begin{equation}
b_0 = 1, \quad b_1 = \frac{1}{2}\sqrt{\frac{\pi}{2}}, \quad b_2 = \frac{5}{12}, \quad b_3 = \frac{15}{64}\sqrt{\frac{\pi}{2}},
\quad b_4 = \frac{221}{1008}.
\end{equation}
Following the same steps as above we find
\begin{eqnarray}
J_2 &= x \sum^{\infty}_{n=0} \frac{(-\lambda)^n}{n!} 2^{n/2} \frac{b_n}{2\sqrt{\pi}} D^{(n-1)/2} \int^L_0 t^{3(n-1)/2}\rmd t\nonumber\\
&= x \sum^{\infty}_{n=0} \frac{(-\lambda)^n}{n!} d_n D^{(n-1)/2} L^{(3n-1)/2}
\end{eqnarray}
where
\begin{equation}
d_n = 2^{n/2} \frac{b_n}{(3n-1)\sqrt{\pi}}.
\end{equation}

Hence we have
\begin{equation}
\av{e^{-\lambda A}} = 1 + x \sum^{\infty}_{n=1} C_n \frac{(-\lambda)^n}{n!} D^{(n-1)/2} L^{(3n-1)/2} + \mathscr{O}(x^2)\label{eq:moments}
\end{equation}
where $C_n = c_n + d_n$ and the first few values are
\begin{equation}
C_1 = 1, \quad C_2 = \frac{32}{15\sqrt{\pi}}, \quad C_3 = \frac{15}{8}, \quad C_4 = \frac{4064}{693\sqrt{\pi}}.
\end{equation}
The first two values are in perfect agreement with those derived in Ref.~\cite{Pruessner2004}, and the authors
are unaware of any previous calculations of $C_n$ for $n > 2$.
Thus we may immediately identify the exponents $\tau = 4/3$ and $\Delta = 3/2$ and the amplitudes allow us to compute
universal amplitude ratios, which we will use later to compare the numerics against theory.

\subsection{Crossover from Branching Process}
The convergence of $\phi_i(x)$ to the normal distribution occurs only as $i \rightarrow \infty$, and hence
the results above are only valid for $L \rightarrow \infty$.
In using the Brownian curve instead of the exact curve described by $P(s;1,m)$ we have taken a hydrodynamic limit
and therefore thrown away any information about the statistics of the process for small $L$.
It is natural, therefore, to ask how we expect the results to differ in this regime.
We propose the existence of an $n$ dependent crossover length, $\xi_n$, such that the above scaling
analysis is valid for $L \gg \xi_n$.
We argue that for smaller systems, $1 \ll L \ll \xi_n$, we expect to see scaling corresponding to the branching process.

Consider adding a particle to the first site.
If the probability that the site has $z$ particles, $p_z$, has support for all $z \in [0,n-1]$, then for $n \gg 1$
it is likely that $0 \ll z \ll n-1$.
In this regime we may assume that the number of times the site topples due to this added particle, which we denote $s_1$,
is largely independent of $z$.
Site $2$ therefore receives $s_1$ particles, each of which may cause it to topple $s_2^j$ times, $i = 1 \ldots s_1$ with
the total number of topplings of site 2, $s_2 = \sum_j s_2^j$.
While $z$ remains far from $0$ and $n-1$, the $s_2^j$ will be largely uncorrelated and by continuing this argument to
more sites, we see that while $s_i$ remain small each site will topple nearly independently.
However, as we continue through the system to higher $i$ the $s_i$ will begin to see large fluctuations and the
avalanches will become correlated, assuming the scaling of the previous section.
Hence, we argue that for systems with $1 \ll L \ll \xi_n$, the avalanches will resemble those of the uncorrelated branching process
with exponents $\tau = 3/2$ and $\Delta = 2$ \cite{Harris1963}.
For larger systems, $L \gg \xi_n$, temporal correlations emerge in the avalanches and $\tau = 4/3$, $\Delta = 3/2$.

The fact that the above argument relies on realisations where $p_z$ has support for a large range of $z$ indicates that the
crossover length $\xi_n$ depends on the details of the toppling rules and as such cannot be thought to have any
``universal'' qualities.
Indeed, we have not specified how the toppling rules in a realisation should be altered as $n$ is increased, and so it is
impossible to say anything {\em a priori} about the behaviour of $\xi_n$.

\section{Numerics}
We now support our claims with numerics by demonstrating that the correct scaling (with crossovers - see previous section) occurs for 
a particular realisation of this directed sandpile model.
In order to study the scaling we choose a realisation such that it is clear how to generalise to higher $n$.
The only remaining difficulty is to find the correct variance to put into the equations when we come to compare with numerics.
In all that follows, we use $2D = \tilde{Q}^{(2)}_{1,\infty}$ as we find that it fits the data very well.

We compare the scaling predicted above with numerics from a realisation with the following toppling rules:
A site $i$, $1 < z_i < n-1$ which receives a particle will topple 1,2 or 3 times with probability $1/8$ or will
not topple with probability $5/8$.
A site with $z_i = 0$ will topple once with probability $3/8$,
a site with $z_i = 1$ will topple once with probability $2/8$ and twice with probability $1/8$ and
a site with $z_i = n-1$ will topple once with probability $6/8$, and 2 or 3 times each with probability
$1/8$.
A site with $z_i = n-1$ has to topple at least once in accordance with restriction (i).

\begin{figure}
\centering
\includegraphics[scale=0.7,clip=true]{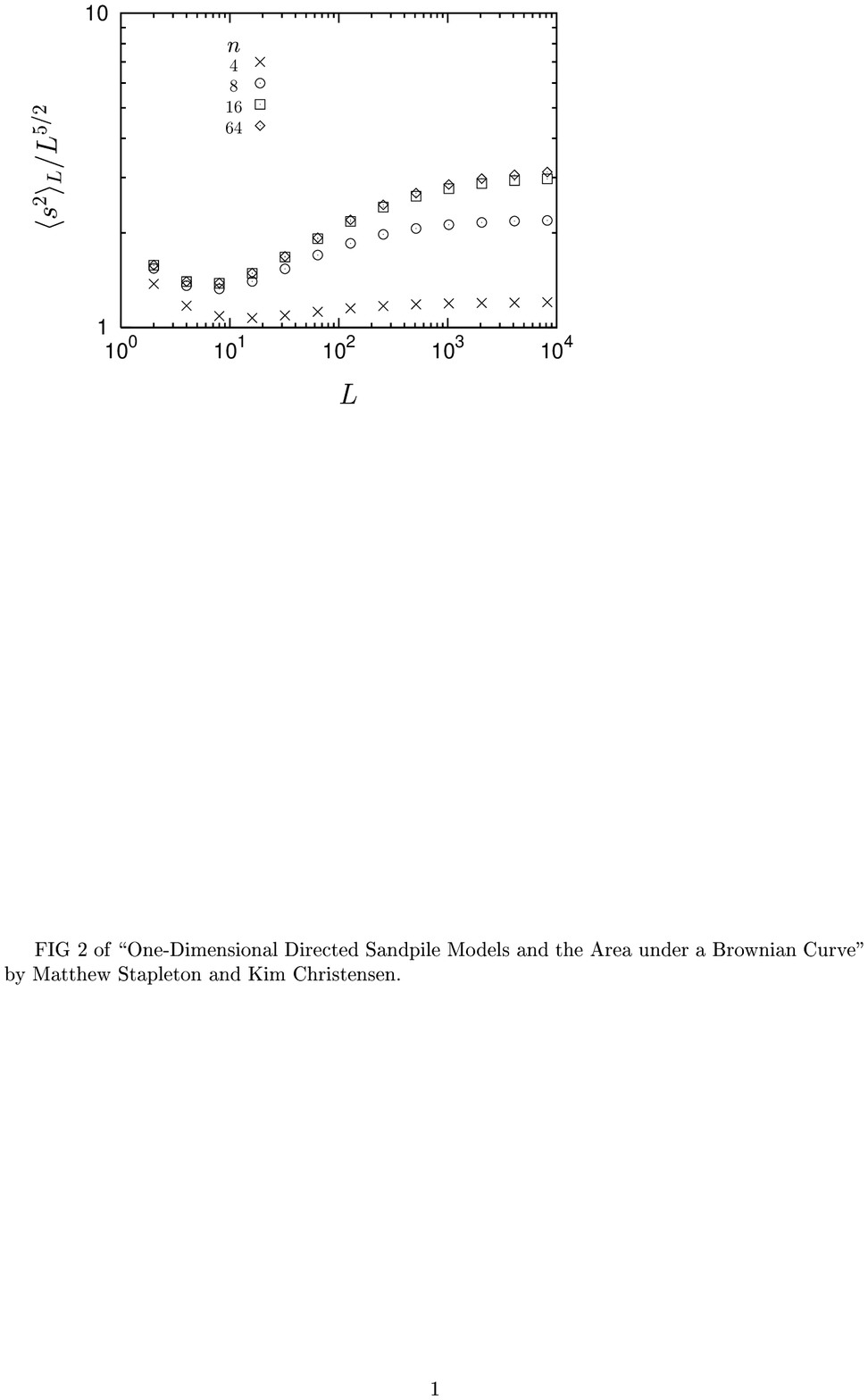}
\caption{Numerical results for $n = 4,8,16,64$.  The errors for both graphs
were calculated using Efron's Jackknife \cite{Efron1982}.
(a) The rescaled second moment $\langle s^2 \rangle_L$ vs. system size.
For large systems this is a constant for all values of $n$.
(b) The moment ratio $g_3(L)$.
For large $L$ this approaches the constant value $g_3 \approx 1.29$ for all values of $n$.
}
\label{fig:numscaling}
\end{figure}

We expect $\langle s^2 \rangle$ to scale with the system size
\begin{equation}
\langle s^2 \rangle_L \sim \begin{cases}L^3 \qquad 1 \ll L \ll \xi_n,\\
L^{5/2} \qquad L \gg \xi_n\end{cases}
\end{equation}
where $\xi_n$ is a correlation length with some (as yet unknown) $n$-dependence.
These results have been confirmed and are shown in figure~\ref{fig:numscaling}.

\begin{figure}
\centering
\includegraphics[scale=0.7,clip=true]{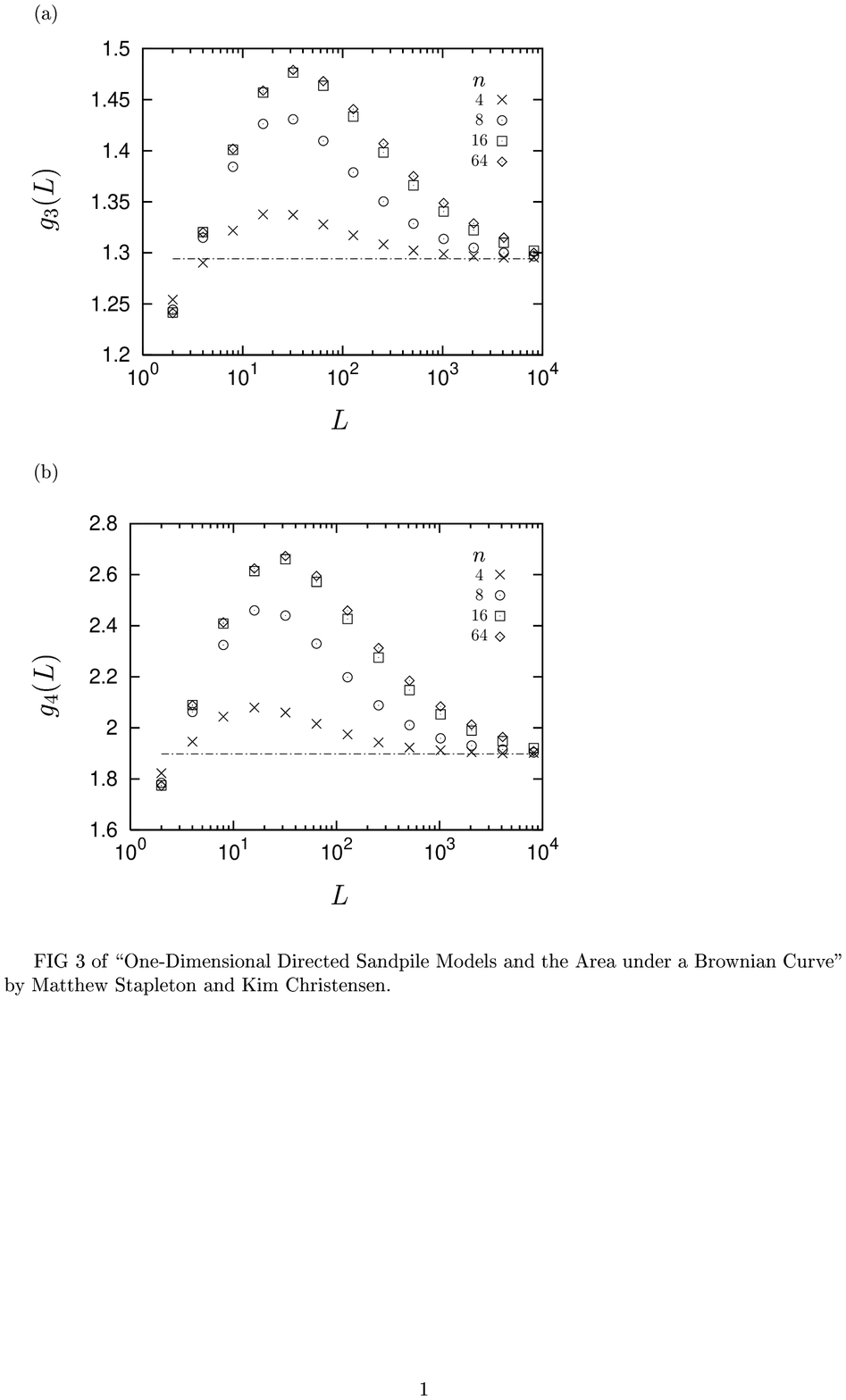}
\caption{Numerical results for $n = 4,8,16,64$.
The errors for both graphs were calculated using Efron's Jackknife \cite{Efron1982} and are approximately the same size as the symbols.
(a) The moment ratio $g_3(L)$ (a) and $g_4(L)$ (b).
For large $L$ these approach the constant values $g_3 \approx 1.29$ and $g_4 \approx 1.9$ respectively for all values of $n$.
The dashed lines indicate the exact values $g_3 = 1.2942\ldots$ and
$g_4 = 1.8975\ldots$, in excellent agreement with the numerics.
}
\label{fig:ratios}
\end{figure}

We also analyse the moment ratios defined by
\begin{equation}
g_k(L) = \frac{\langle s^k \rangle_L \langle s \rangle_L^{k-2}}{\langle s^2 \rangle^{k-1}_L}.
\end{equation}
It is a straightforward calculation to show that, for an avalanche probability given by
\eref{eq:simplescaling} $g_k(L)$ approach universal values for $L \rightarrow \infty$.
These values are simply ratios of the amplitudes $C_n$ calculated in \ref{sec:moments},
\begin{eqnarray}
g_k  &\equiv \lim_{L \rightarrow \infty} g_k(L)\nonumber\\
&= \frac{C_k}{C_2^{k-1}}
\end{eqnarray}
This agrees with the numerics, as illustrated in figure~\ref{fig:ratios} for $g_3(L)$ which
appears to converge to a universal value of $g_3(\infty) \approx 1.29$, in excellent agreement with the theoretical
prediction $g_3 = 15^3\pi / 2^{13} = 1.294\ldots$ as well as numerics for a different realisation
published elsewhere \cite{Stapleton2005}.
This supports our claim that the $L \rightarrow \infty$ limits of the $g_k(L)$ are indeed universal.
Note also that $\xi_n$ has a notably different $n$ dependence in this model than in the one presented in
\cite{Stapleton2005}.
In this case, $\xi_n$ saturates to a constant value for large $n$, meaning that for large $n$ the crossover
occurs at the same value of $L$.
This is because the support of $p_z$ is finite for $n \rightarrow \infty$.

\section{Conclusion}
We have found the stationary state avalanche-size distribution for a general $n$-state directed sandpile
model.
The avalanches can be mapped onto a random walk of dependent random variables and, using an applicable central limit
theorem, we have shown that under a broad set of conditions the moments scale with $\tau=4/3$ and ${\Delta}=3/2$.
We also note that this value of $\tau$ agrees precisely with that obtained in \cite{Kearney2005}, which calculates the probability distribution
in the infinite system size limit.
We have also calculated the moment generating function for the area under a random walker with an absorbing boundary, and
found a relation for the moment amplitudes in terms of those already known for other Brownian processes.

\ack{The authors wish to thank G. Pruessner and D. Dhar for discussions and B. Derrida for help in finding the area under the random walker.
M.S. gratefully acknowledges the financial support of U.K. EPSRC through grant No. GR/P01625/01.}

\appendix

\section{Explicit Example: $n=2$.}
\label{sec:precise}
In this section we calculate the steady state properties of an $n=2$ model, which is a generalisation
of the model studied in Ref.~\cite{Pruessner2004}, and compare predictions to numerical simulation.
For $n=2$, the most general model we can write down, which obeys the rules (i)-(iv), is
\numparts
\begin{eqnarray}
\op{S}_0 &= \left(\begin{array}{ccc} 0 & 0 \\ \alpha & 0 \end{array}\right),\\
\op{S}_1 &= \left(\begin{array}{ccc} 1-\alpha & 0 \\ 0 & 1-\beta \end{array}\right),\\
\op{S}_2 &= \left( \begin{array}{ccc} 0 & \beta \\ 0 & 0 \end{array}\right),
\end{eqnarray}
\endnumparts
where $0 \leq \alpha \leq 1$ is the probability that a site with $z = 0$ does not topple on receiving a particle and
$0 \leq \beta \leq 1$ is the probability that a site with $z=1$ topples twice on receiving a particle.
Hence, the single site toppling matrix is
\begin{equation}
\op{G}_1(x) = 
\left( \begin{array}{ccc} x - \alpha x & \beta x^2\\[0.5cm] \alpha &  x - \beta x \end{array}\right).
\end{equation}
We now proceed to calculate the steady state properties of this model, following the prescription
given in \ref{sec:stationary}.
$\op{G}_1(x)$ has eigenvalues $\lambda = x$, $\mu = x(1 - \alpha - \beta)$ and eigenvectors
\numparts
\begin{eqnarray}
\ket{e_{\lambda}(x)} &= \frac{1}{\alpha + \beta}\left( \begin{array}{c}\beta x\\ \alpha\end{array}\right), \\
\quad \bra{e_{\lambda}(x)} &= \left( \begin{array}{c}\frac{1}{x},1\end{array}\right),\\
\ket{e_{\mu}(x)} &= \left( \begin{array}{c}-x\\1\end{array}\right),\\
\quad\bra{e_{\mu}(x)} &= \frac{1}{\alpha + \beta}\left( \begin{array}{c}\frac{-\alpha}{x},\beta\end{array}\right).
\end{eqnarray}
\endnumparts
Hence, the eigenvector for the stationary state is
\begin{equation}
\ket{0}_L = \ket{e_{\lambda}(0)}^{\otimes L} = \left(\frac{1}{\alpha + \beta} \left( \begin{array}{c} \beta \\ \alpha \end{array}\right)\right)^{\otimes L}
\end{equation}
valid for $| \mu | \neq 1$.

\begin{figure}
\centering
\includegraphics[scale=0.7,clip=true]{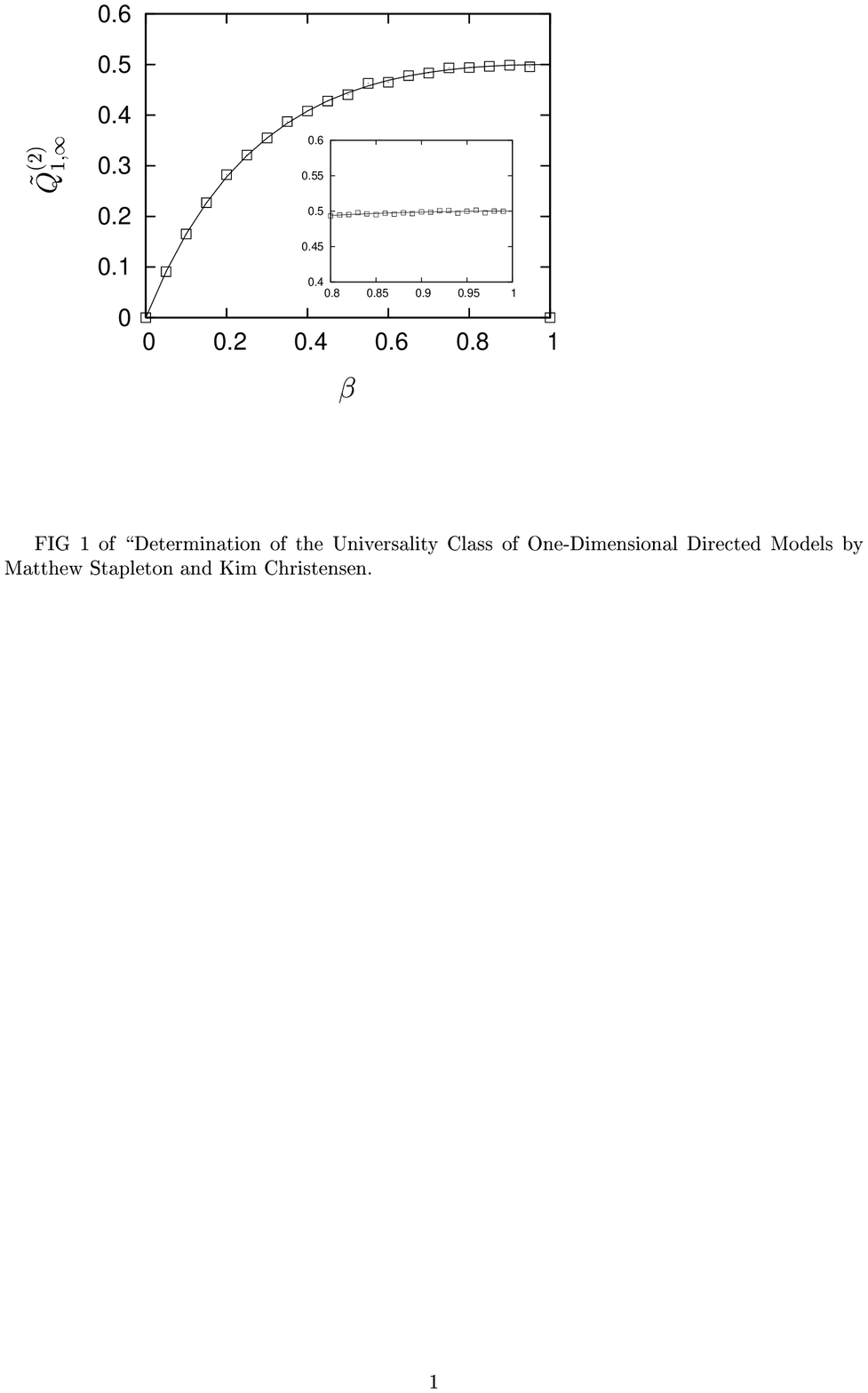}
\caption{
Numerical results for $\tilde{Q}^{(2)}_{1,\infty}$ for $\alpha = 1$ as a function of $\beta$ with data (squares) compared
against the values predicted by \eref{eq:qtilde} (solid line).
The data was obtained by measuring $\tilde{Q}^{(2)}_{1,m}$ for large $m$ and estimating the asymptotic value.
Comparison is made across the whole range of $\beta$ and
the inset shows data in the vicinity $\beta \rightarrow 1$.
Note that the agreement is excellent right up to $\beta = 1$.
Typical error bars for the numerical data are the size of the squares.}
\label{fig:Q_1_inf}
\end{figure}

\begin{figure}
\centering
\includegraphics[scale=0.7,clip=true]{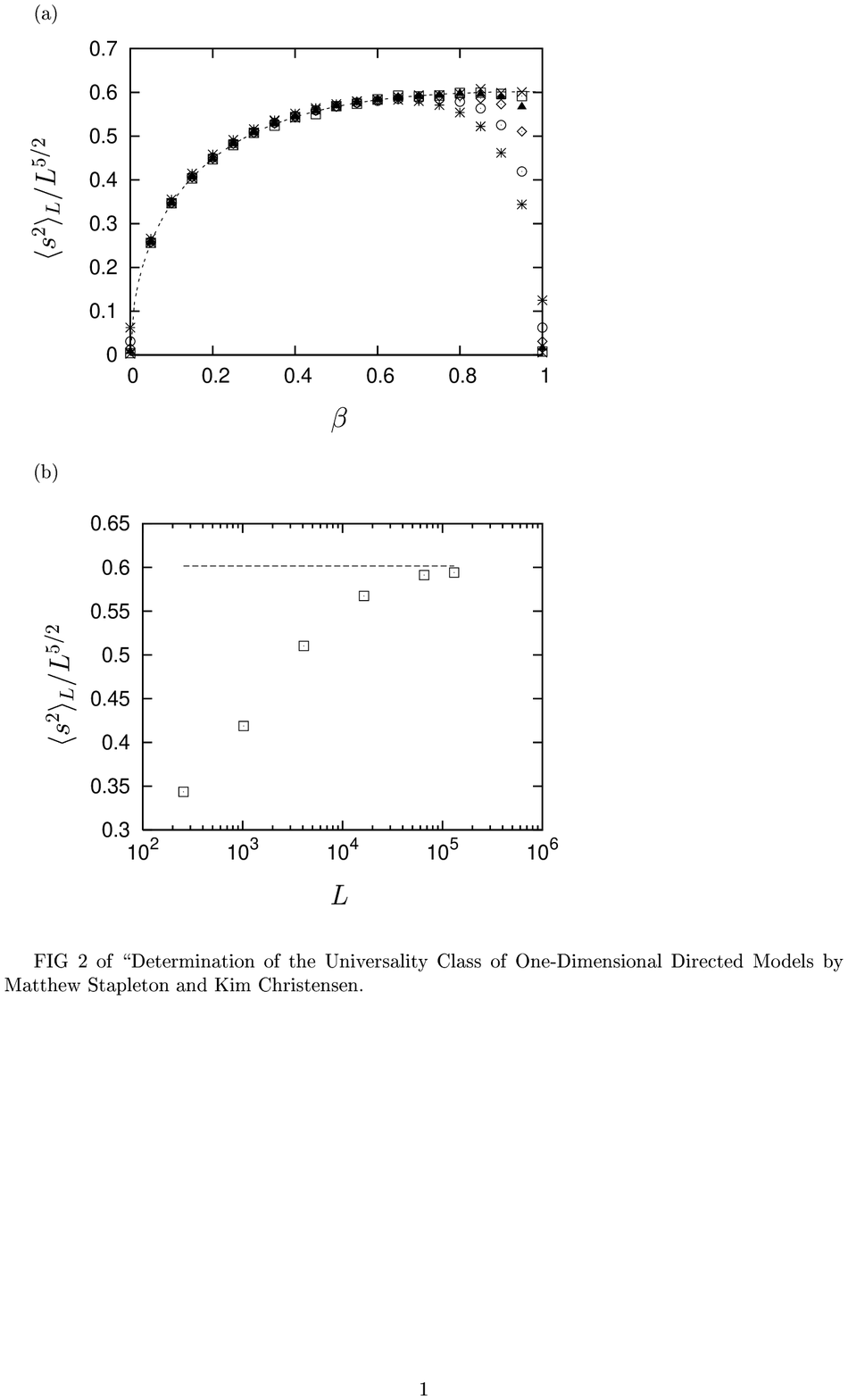}
\caption{
Numerical results for the rescaled second moment $\mom{s}{2}_L / L^{5/2}$ for $\alpha = 1$
(a) as a function of $\beta$ for $L=256$ ($*$), 1024 ($\circ$), 4096 ($\diamond$), 16385 ($\blacktriangle$), 65536 ($\square$) and
131072 ($\times$).
(b) Rescaled second moment $\mom{s}{2}_L / L^{5/2}$ for $\beta = 0.95$ vs inverse system size.
The dashed line is the theoretical value.
The measurements appear to converge towards the theoretical line large $L$, supporting our claim that the deviation is a finite size effect.}
\label{fig:Manna}
\end{figure}

From these results it follows immediately
\begin{equation}
\tilde{Q}^{(2)}_{1,\infty} = p_0 p_1 + p_1 p_0 = 2 \frac{\alpha \beta}{\alpha + \beta}\label{eq:qtilde}
\end{equation}
and hence, from \eref{eq:moments} and using $2D = \tilde{Q}^{(2)}_{1,\infty}$,
\begin{equation}
\mom{s}{2}_L \sim \frac{32}{15(\alpha + \beta)}\sqrt{\frac{\alpha\beta}{\pi}} L^{5/2},
\end{equation}
in perfect agreement with numerics, see figure~\ref{fig:Q_1_inf} and figure~\ref{fig:Manna}.
However, it should be noted that for $\alpha$ and $\beta$ both approaching 1 the random walker it describes will spend
more and more time on either only odd or only even sites.
Hence, it will take longer times (larger system sizes) for the statistics to reach the asymptotic values and so
we expect very strong corrections to scaling for $\alpha, \beta \rightarrow 1$.
When $\alpha = \beta = 1$, we no longer have a unique stationary state and so scaling is not observed.

\section{Calculating the Amplitudes $C_n$}
The amplitudes $a_n$ and $b_n$ appearing in \ref{sec:moments} can be calculated using the methods outlined in Refs.~\cite{Majumdar,Perman1996}.
For $a_n$ we define
\begin{equation}
a_n = (\sqrt{2})^{-n} \frac{\Gamma(1/2) n!}{\Gamma(\frac{3n+1}{2})} R_n
\end{equation}
where $R_n$ are constructed through the following recursion relations
\begin{eqnarray}
R_n &= \beta_n - \sum^n_{j=1} \gamma_j R_{n-j}\\
\beta_n &\equiv \gamma_n + \frac{3}{4}(2n-1)\beta_{n-1}\\
\gamma_n &\equiv \frac{\Gamma(3n+1/2)}{\Gamma(n+1/2)}\frac{1}{(36)^n n!}.
\end{eqnarray}
Similarly, for the $b_n$
\begin{eqnarray}
b_n &= 4 (\sqrt{2})^{-n} \frac{\Gamma(1/2) n!}{\Gamma(\frac{3n-1}{2})} K_n\\
K_n &\equiv \frac{3n-4}{4} K_{n-1} + \sum^{n-1}_{j=1} K_j K_{n-j}.
\end{eqnarray}
Putting these together and rearranging slightly, we find that the amplitudes $C_n$ are given by
\begin{equation}
C_n = \frac{n!}{\Gamma(\frac{3n+1}{2})} \left( R_n + 2K_n \right)
\end{equation}
We tabulate the first 10 values of $C_n$, along with the universal amplitude ratios $g_n = C_n / C_2^{n-1}$ in 
table B1.
\newpage
\begin{table}
\caption{Tabulated values of $C_n$ and $g_n$.}
\begin{indented}
\item[] \begin{tabular}{@{}lll}
\br n & $C_n$ & $g_n$\\
\mr 1 & 1 & 1\\[0.5cm]
2 & $\frac{32}{15 \sqrt{\pi}}$ & 1\\[0.5cm]
3 & $\frac{15}{8}$ & $\frac{3375}{8192}\pi$\\[0.5cm]
4 & $\frac{4096}{693\sqrt{\pi}}$ & $\frac{47625}{78848}\pi$\\[0.5cm]
5 & $\frac{2875}{448}$ & $\frac{145546875}{469762048}\pi^2$\\[0.5cm]
6 & $\frac{1219336}{51051\sqrt{\pi}}$ & $\frac{38580553125}{71374471168}\pi^2$\\[0.5cm]
7 & $\frac{745039}{24576}$ & $\frac{2828819953125}{8796093022208}\pi^3$\\[0.5cm]
8 & $\frac{25796624240}{200783583\sqrt{\pi}}$ & $\frac{10202766423046875}{15969609677012992}\pi^3$\\[0.5cm]
9 & $\frac{214422265}{1171456}$ & $\frac{549540812759765625}{1288029493427961856}\pi^4$\\[0.5cm]
10 & $\frac{15033906553126}{17468171721\sqrt{\pi}}$ & $\frac{3567616496493767578125}{3793868231748622483456}\pi^4$\\
\br
\end{tabular}
\end{indented}
\label{tab:values}
\end{table}

\end{document}